\documentstyle[aps,prl,epsf,floats,twocolumn]{revtex}
\draft
\begin{document}

\twocolumn[\hsize\textwidth\columnwidth\hsize\csname @twocolumnfalse\endcsname
\title{Negative Electron Drag and Hole-Like behavior in the Integer
Quantum Hall Regime}
\author{X. G. Feng,  S. Zelakiewicz, H. Noh, T. J. Ragucci, and T. J. Gramila}
\address{Department of Physics, The Pennsylvania State University, 
University Park, PA 16802}
\author{L. N. Pfeiffer and K. W. West}
\address{Bell Labs, Lucent Technologies, Murray Hill, NJ 07974}
\date{\today}

\maketitle

\begin{abstract}
Electron drag between two two-dimensional electron gases in magnetic fields
has been observed with a polarity opposite that for zero field.  This
negative drag requires that the electrons have a hole-like dispersion.
Density dependence measurements in the integer quantum Hall regime show
that drag is negative only when the upper Landau level of one layer is more
than half filled while the other is less than half filled, indicating that
hole-like dispersion is present in a half of each Landau level.  Negative
drag is argued to be a consequence of disorder.
\end {abstract}
] 

The integer quantum Hall effect (IQHE) is a central element of
two-dimensional electron gas (2DEG) physics, in which disorder plays a
critical role.  Numerous experimental techniques\cite{Book} have
explored aspects of the effect.   A new approach, electron
drag, has recently been applied to the study of this important topic.  This
approach, in which electron-electron (e-e) scattering between two
parallel 2DEGs is measured, can provide information complementary to
single layer measurements.  In particular, it was predicted\cite{Hu}
and observed\cite {Hill,Klitzing} that for each individual
magneto-resistance peak, a double-peaked structure would exist for
drag with matched density electron layers, due to an enhanced
importance of screening.  In this Letter, drag measurements exploring
the case of unmatched densities in detail are presented, which reveal
a new regime of electron drag.  In this regime, the polarity of the
measured drag voltage is {\it opposite} that normally
found\cite{Hill,Klitzing,Gramila,Solomon,Gramila2,Hill2,Noh} for
electron layers, matching instead the polarity of an electron-hole
(e-h) system\cite{Sivan}.  The polarity of the drag signal provides a
new and direct experimental observation about the nature of the states
in the IQHE; that a hole-like dispersion exists in one half of each
Landau level (LL).  We argue, furthermore, that disorder is a crucial
element in this hole-like behavior.

In electron drag, when a current is driven through the first of two
electrically isolated 2DEG's, inter-layer e-e scattering transfers
momentum to the second (drag) layer electrons.  If no current is allowed
to flow from the drag layer, a voltage develops from charge accumulation
to precisely cancel the effective inter-layer drag force.  The ratio of
this drag voltage to the drive layer current, the drag transresistance
$\rho_D$, has been shown\cite{Gramila} to directly determine the
inter-layer e-e scattering rate.  Of particular interest to this work is
the drag voltage polarity.  In zero field, the polarity is {\em
opposite}\cite{Gramila} that of the drive layer voltage, as it results
from charge accumulation in the direction of the drive layer drift
velocity.  For an e-h system, by contrast, the drag polarity is
the {\em same}\cite{Sivan} as the drive layer voltage. (We refer to the
drag polarity of the e-e/e-h case as positive/negative).  Previous drag
experiments\cite{Hill,Klitzing} between electron layers with matched
densities find a {\em positive} drag throughout the IQHE regime.  Our
measurements, in which electron densities were systematically varied by
application of an interlayer bias and/or an overall top gate voltage,
showed both positive and negative drag.

\begin{figure}[!b]
\begin{center}
\leavevmode
\hbox{%
\epsfysize = 3in
\epsffile{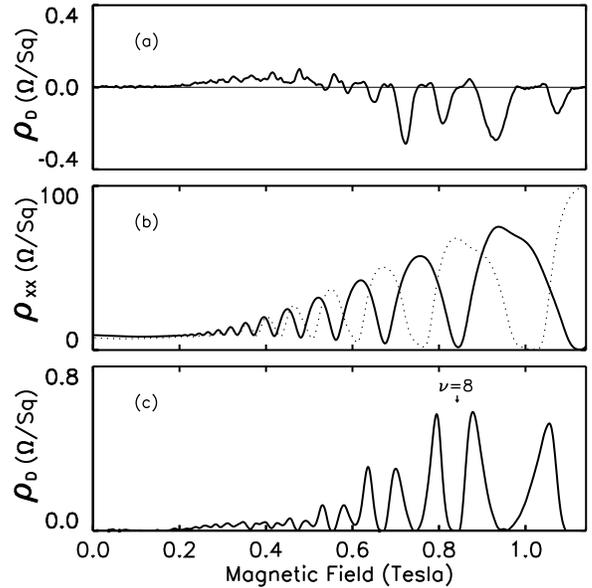}}
\end{center}
\caption{ Drag versus magnetic field for matched (c) and unmatched (a)
        layer densities at 1.15K.  The longitudinal magneto-resistance (b)
        for the drive layer in both cases ($1.66\times 10^{11}/cm^{2}$)
        (solid line), the dotted line is the drag layer with a mismatched
        density ($1.49\times 10^{11}/cm^{2}$).  A clear difference in
        filling factors is evident for negative drag (a).  }
\label{fig1}
\end{figure}

The sample was a GaAs/AlGaAs double quantum well structure with two
$200$\AA\/ wells separated by a $225$\AA\/ barrier, each with an
electron density of $\sim1.6\times 10^{11}/cm^{2}$ as grown, and
mobilities of $2.5\times 10^{6}cm^{2}/Vs$.  The striking implications of
negative drag require we ensure the observed polarity is not due to
well-known complications of transport in magnetic fields or other drag
difficulties.  Standard tests\cite{Gramila} of changing drag layer
ground to test for inter-layer leakage, varying drive  current to
check for linearity, and interchanging drive and drag layers, all
confirm the measurements validity.  In addition, reversing the magnetic
field yields no change in the signals, indicating that the drag is
longitudinal, i.e., the Hall voltages which develop in these fields
play no role.

Negative drag emerges when the densities of the electron layers are {\it
not} matched.  A comparison of $\rho_D$ for matched and unmatched
densities is shown in Fig.~1.  For matched densities, each layer has a
magneto-resistance ($\rho_{xx}$) similar to the solid curve of
Fig.~1b, with the corresponding drag versus field shown below.  The
measurements are comparable to those previously showing the
double-peak structure \cite{Hill,Klitzing}.  If the the drag layer
density is decreased 10\%, however, with $\rho_{xx}$ shown by the
dotted line in Fig.~1b, the negative drag of Fig.~1a is obtained.

The existence of negative drag, the same  polarity as the the e-h
system\cite{Sivan}, has implications for the nature of the states in the
IQHE regime.  It would appear to suggest holes have a role in the
process, but this is clearly inconsistent with the sign of the Hall
voltage.  Instead, negative drag with two electron layers requires interlayer
scattering induce a drift of drag layer electrons in a direction {\it
opposite} the drive layer drift velocity, contrary to expectations of
classical momentum conservation.  Such expectations, however, are
complicated by the presence of a magnetic field, $B$.  In the Landau gauge,
in particular, where states are infinite strips along the current
direction $y$, it is the canonical momentum $\hbar k_y$ which is
conserved during scattering, and not the classical momentum or velocity.
To obtain a negative drag signal, a scattering event which
transfers $\Delta k_y$ from one layer to the other must result in a
change in velocity, $dE/dk_y$, in the {\it same} direction in both
layers.  Negative drag thus places a strong constraint on the electron
dispersion; $d^2E/dk_y^2$ must have different signs in the two layers.

The dispersion relation required for negative drag should not exist for 2-D
electrons in magnetic fields in the absence of disorder.  For a system with
hard-wall confinement, the $E$-$k_y$ dispersion for a single LL has been
shown\cite{Heuser} to be flat when the electron center of mass is near the
middle of the sample, with $E$ increasing as the electron comes within a
cyclotron radius of the sample edge.  Although electrons on both sides of
the sample have opposite velocities, the change in velocity with $k_y$,
$d^2E/dk_y^2$, is always positive or zero.  The absence of a `hole-like'
dispersion (i.e. $d^2E/dk_y^2 < 0$) in this system implies that
negative drag is a disorder-related effect.

It is widely recognized that disorder broadening of Landau levels is an
important element of the IQHE.  A significant complication introduced
by disorder, though, is that $k_y$ is no longer strictly a good quantum
number, as disorder lacks translational symmetry. It still remains possible
to consider properties of the dispersion relation.  In
particular, the use of periodic boundary conditions, as in the numerical
studies of Ref. \cite{Ohtsuki1,Ohtsuki2,Viehweger}, preserve the Bloch
momentum along $y$ as a conserved quantity.  (In such cases, only the
change in $k_y$ is physically meaningful).  Qualitative information
about the dispersion relation deduced from our measurements is valid if
we consider the lack of true translational symmetry in a similar manner.

\begin{figure}[!b]
\begin{center}
\leavevmode
\hbox{%
\epsfysize = 2.45in
\epsffile{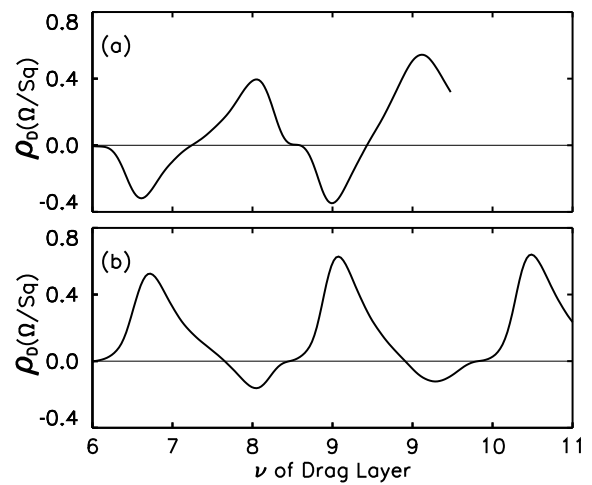}}
\end{center}
\caption{Drag versus drag layer filling factor(density).  The drive
	layer density is fixed at $1.66\times 10^{11}/cm^{2}$, with
	magnetic fields for (a)/(b) of 0.92/0.81 Tesla, and drive layer
	$\nu$ of 7.39/8.47. Nearly periodic behavior is observed.}
\label{fig2}
\end{figure}

Ascertaining the conditions required for negative drag is clearly
important.  A rough correspondence in $B$ between the transition from
positive to negative drag (Fig.~1a) and the onset of significant
differences in the Schubnikov-de Haas oscillations (Fig.~1b), suggests that
relative LL filling plays a role.  To test this, $\rho_D$ was measured at
fixed $B$ and drive layer density while varying the drag layer density, or
equivalently its filling factor $\nu$.  Odd integer $\nu$ corresponds to
two half filled LL's, as spin is not resolved.  As shown in Fig.~2a, with a
drive layer LL more than half filled ($\nu = 7.39$), $\rho_D$ shows
alternating positive and negative peaks.  Drag is positive when the highest
drag layer LL is more than half filled; negative drag appears only when
that LL is less than half filled, irrespective of the overall LL index.
Changing $B$ so the upper drive layer LL is less than half filled ($\nu =
8.47$), (Fig.~2b) reverses the polarity of drag versus drag layer $\nu$.
The symmetry in this behavior clearly establishes a negative drag criteria:
the deviation from half filling must be opposite in the two layers.  The
observation that the drag polarity is nearly periodic in the drag layer
$\nu$ shows further that the sign of $d^2E/dk^2$ is directly related to the
position of the electron states within each LL, with hole-like dispersion
existing for each Landau level.

The drag temperature dependence provides further information about the
dispersion.  While the density dependence measurements require a different
dispersion for states above or below half-filling, it provides little
information regarding energy differences of the states.  Measurements
exploring such differences are shown in Fig.~3, which follows the evolution
of drag for 10\% mismatched densities as temperatures range from 1.15 K to
2.18 K.  The most striking element of these data is the decrease of
negative drag with increasing temperature, disappearing altogether at
roughly 1.8 K.  Above this temperature, drag becomes positive and
continues to increase. 

\begin{figure}[!t]
\begin{center}
\leavevmode
\hbox{%
\epsfysize = 2.45in
\epsffile{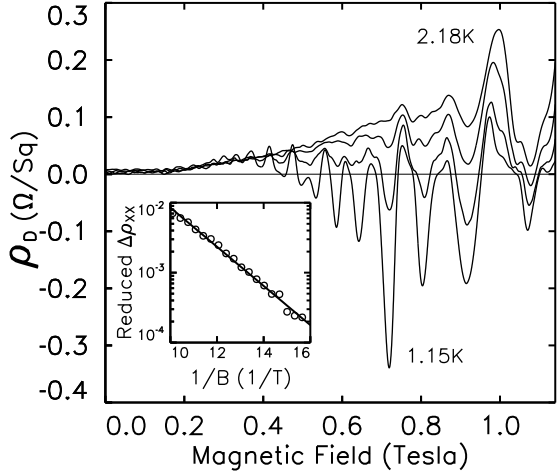}}
\end{center}
\caption{The temperature dependence of negative drag, showing its
	 disappearance above $\sim$1.8K.  Densities of the two layers
	 are 1.66 and 1.49$\times 10^{11}/cm^{2}$; temperatures of
	 1.15K, 1.53K, 1.87K, and 2.18K are shown.  Inset: A Dingle plot
	 of the reduced oscillations in $\rho_{xx}$ at low fields, used
	 to determine a Landau Level energy width of 1.7K.}
\label{fig3}
\end{figure}

These data gauge the energy difference between the electron and hole-like
states within the Landau levels.  If electrons in both layers have the same
character dispersion, electron or hole-like, their scattering contributes
to positive drag.  Negative scattering requires a different character
dispersion in each layer.  While the states which contribute to the net
interlayer scattering are determined in large part by the level of the
Fermi energy in the LL, raising the temperature permits participation of
states further from this energy.  Once the temperature becomes
comparable to the energy difference between the electron and hole-like
states, the contributions of both processes are present and will tend to
cancel.  The determination that negative drag uniformly disappears near
1.8K indicates, then, that the energy difference between states of
differing dispersion is approximately this temperature.

This characteristic energy can provide a test for determining the physical
origin of the hole-like dispersion.  Unfortunately, no theoretical
investigations with scattering calculations from hole-like states in the
IQHE regime yet exist for comparison.  Indeed, the most extensive
theoretical investigations of drag in this regime\cite{Bonsager} explore no
physics which would result in a negative drag signal.  Two possible origins
of this behavior should be considered and compared to measurement
regardless of this absence.  While the two sources are fundamentally
different, disorder plays a central role in each.  In the first, negative
drag arises purely from the bulk states of the LL, while in the second, it
arises from Quantum Hall edge states.

While LL bulk states show no dispersion in the absence of disorder, its
presence generates an energy width for the states, which could result in
hole-like dispersion.  Motivation for such a scenario arises from
consideration of states in the Landau gauge in which the impurity
potential is included (as opposed to reduced to a scattering potential),
resulting in a set of states in which the local electron energy varies
with the local potential.  In a slice across the sample, the energy
would vary above and below the average cyclotron energy.  Since the
position across the sample is related to $k_y$ in this gauge, there
could exist both electron and hole-like dispersion. 

Consideration of a bulk state source for the hole-like behavior leads to an
energy scale for the disappearance of negative drag, even without a
detailed picture for the dispersion.  The energy width of the bulk states
can be estimated through analysis of Schubnikov-de Haas oscillations at low
temperatures.  Measurements comparing the oscillation amplitude, $\Delta
\rho_{xx}$, to the inverse field at .5K is inset in Fig.~3.  The
oscillations are shown in the reduced form (a Dingle
plot)\cite{Dingle,Fletcher} used for determination of the quantum lifetime
$\tau_0$: $\Delta\rho_{xx}/2D(X)\rho_o$, where $D(X) = X/sinh(X)$ with $X =
2\pi^2k_bT/\hbar\omega_c$, and with $\omega_c$ and $\rho_o$ the cyclotron
frequency and zero field $\rho$, respectively.  The slope of a linear fit
to the data (solid line) determines $\tau_0$, yielding a disorder induced
energy width, $\hbar/\tau_o$, of 1.7K.  All states in the LL should be
accessible and thus participate in scattering once the temperature is of
this order.  The observation that negative drag disappears at roughly 1.8K
thus provides support for a bulk state source for the hole-like dispersion.

The second source considered involves edge states.  The presence of
disorder can permit edge states of lower LL's to mix with bulk states of a
higher LL.  States with energies well below the bulk state LL will be
little influenced, but as the energy approaches that of the bulk states,
the resultant mixed states will acquire greater bulk-like character.  Above
half filling, the transition is from bulk to edge-like character.  The
inset of Fig.~4a compares
a schematic of the resultant energy level diagram (top)  to
the case in the absence of disorder (bottom).
Such mixing has been explored in numerical
investigations\cite{Ohtsuki1,Ohtsuki2,Viehweger}, our simple diagram is
drawn from that work.  The theoretical work addresses a key question regarding
the dispersion: how $k_y$ varies within these states.  It increases as the
curves are traversed in a clockwise fashion, as shown by arrows in the
figure.  The mixing of bulk and edge states thus provides a mechanism for
hole-like dispersion below half filling (i.e. for these states, $d^2E/dk_y^2
< 0$), while states above half filling retain an electron-like dispersion.
Energy differences between electron and hole-like states for this source
would likely be comparable to that determined in Fig.~3, though questions
concerning the relatively small number of edge states, whose influence must
dominate drag measurements, must be considered.

A potential test for the influence of edge states arises through severe
reduction of their number.  Measurements made at 10.1, 6.23, and 4.76
Tesla, and 1.2K are shown in Fig.~4a, b, and c, respectively.  In Fig.~4a,
the drive layer density is fixed so $\nu=0.8$, with the drag layer $\nu$
varied between 0.4 and 0.8.  Above $\nu \sim 0.5$, i.e. with both layers'
LL's more than half filled, drag is positive, as expected.  Reducing the
drag layer $\nu$ below 0.5, so the layers have opposing deviations from
half filling, which would yield negative drag at low fields, results
instead in a positive signal.  This measurement remains entirely in the
lowest LL, so no edge states from lower LL's exist.  Similar behavior is
observed in the next LL, where only a single edge state in present.  In
Fig.~4b, with a drive layer $\nu=1.3$, drag remains positive irrespective
of the deviation from half filling, with a clear zero near $\nu=1$.
Remaining within the second LL continues to generate only positive drag; in
Fig.~4c, the drive layer $\nu$ is 1.7 with the drag layer ranging from 1.2
to 1.7.  The observation that negative drag disappears at this temperature
as the number of edge states is greatly reduced appears to support the
involvement of edge states in negative drag in negative drag.  While the
behavior is inconsistent with the simple picture previously considered for
a bulk state source, as LL widths have been measured\cite{jpe_sqrt} to
increase like $B^{0.5}$ at high fields, it remains possible that other
influences of high magnetic fields, such as the enhanced importance of
intra-layer e-e interactions, could instead cause the observed
disappearance of negative drag.

\begin{figure}[!t]
\begin{center}
\leavevmode
\hbox{%
\epsfysize = 2.45in
\epsffile{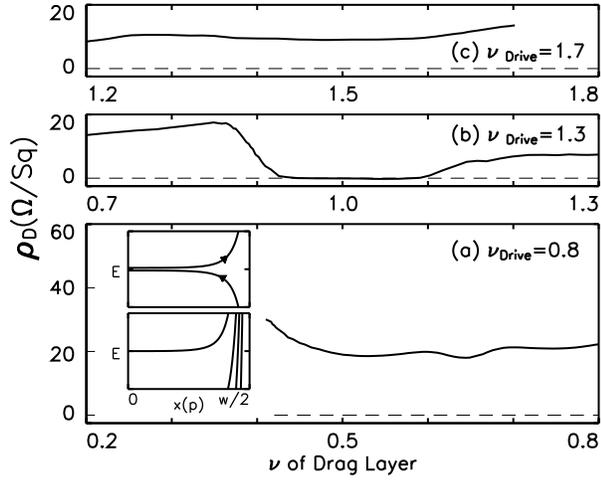}}
\end{center}
\caption{Drag for fields of 10.1 (a), 6.23 (b), and 4.76T (c) at 1.2K vs.
	drag layer $\nu$, with drive layer $\nu$ of 0.8, 1.3, and 1.7
	respectively.  Negative drag is absent in the extreme quantum
	limit.  Inset top: the two hypothetical branches of the mixed
	states near the Landau band, exhibiting hole-like dispersion below
	half filling. Below are the states without disorder.}
\label{fig4}
\end{figure}

Although a full understanding of the effect remains elusive, the
observation of hole-like behavior for electrons in magnetic fields has
revealed, we believe, direct new information about the nature of the states
in the IQHE regime.  Because its observation depends critically on the
density difference between electron layers, it is unlikely the information
is accessible in single layer transport measurements.  The quantitative
results presented here are suitable for comparison to calculation, and may
help further illuminate the important role of disorder in the IQHE.
Negative drag may also be important for analysis of the double peaked
structure seen\cite{Hill,Klitzing} (also Fig.~1c) in drag at matched
densities, argued\cite{Hu} to result from a competition between an
increasing DOS and enhanced screening.  The degree to which the existence
of both positive and negative drag scattering processes contributes to this
effect will require further investigation.

In summary, negative drag has been observed for 2DEG's with mismatched
densities in the IQHE regime. The polarity observed is opposite that for
zero field, and the same as for the electron-hole double layer system.  The
sign the drag polarity exhibits nearly periodic behavior as the filling
factor of one layer is varied, supporting a criteria for negative drag:
if the upper Landau level of one layer is more than half filled, that of
the other must be less than half filled.  The existence of the hole-like
dispersion relation required for negative drag, $d^2E/dk_y^2 < 0$, is
argued to result from disorder.

We appreciate discussions with A.~H.~MacDonald and N. Bonesteel, and the
contributions of J.~P.~Eisenstein to both this work and the drag technique.
Support from NSF grant DMR-9503080, the Sloan Foundation, and the Research
Corp. is gratefully acknowledged.

\end{document}